\documentclass[twocolumn,showpacs,prb]{revtex4-1}

\usepackage[cp1251]{inputenc}
\usepackage[russian]{babel}

\usepackage{graphicx}
\usepackage{dcolumn}
\usepackage{bm}

\begin{document}

\title{Chiral $d$-wave superconductor
with nonzero center-of-mass pair momentum}

\author{V.I. Belyavsky, V.V. Kapaev, and Yu.V. Kopaev}

\affiliation{P. N. Lebedev Physical Institute of Russian Academy of Sciences, Moscow 119991, Russia}%


\begin{abstract}
Hypothetical topologically nontrivial superconducting state of
two-dimensional electron system is discussed in connection with
the problem of pairing with large center-of-mass pair momentum
under predominant repulsive screened Coulomb interaction. Direct
numerical solution of the self-consistency equation exhibits two
nearly degenerate order parameters which can be formally referred
to $d^{}_{x^2-y^2}$ and $d^{}_{xy}$ orbital symmetry. Spontaneous
breaking of the time-reversal symmetry can mix these states and
form fully gapped chiral $d+id$ superconducting state.
\end{abstract}


\maketitle

In recent years, condensed matter physics has significantly
focused on studies of peculiar states of matter, such as
two-dimensional (2D) topological insulators and
superconductors.\cite{Qi_Zhang} Both time-reversal (TR) invariant
and TR breaking topological superconductors have attracted a lot
of interest, in particular, because of their potential
applications. TR breaking superconductors are classified by an
integer topological invariants\cite{Volovik} similar to those used
for classification of quantum Hall states.\cite{Thouless} The
simplest chiral topological triplet superconductor with
$p^{}_x+ip^{}_y$ orbital symmetry was considered by Read and
Green\cite{Read_Green} and predicted to exist in
${\text{Sr}}_2{\text{RuO}}_4$ by Mackenzie and
Maeno.\cite{Mackenzie}

Degeneration of the $d^{}_{x^2-y^2}$ and $d^{}_{xy}$ ordered
states inherent in doped graphene monolayer has recently
considered by Nandkishore et al.\cite{Nandkishore} as possible
origin of a rise of a singlet chiral superconducting (SC) state
with $d+id$ orbital symmetry. Such complex order parameter was
suggested by Laughlin\cite{Laughlin} to connect the TR broken
symmetry and the low-temperature phase transition observed in
${\text{Bi}}_2{\text{Sr}}_2{\text{CaCu}}_2{\text{O}}_8$ in
external magnetic field.\cite{Krishana} In such a case, the
$d^{}_{xy}$ component of the SC order parameter turns out to be
field-induced. Similar phase transition was observed in
${\text{Ni}}$-doped
${\text{Bi}}_2{\text{Sr}}_2{\text{CaCu}}_2{\text{O}}_8$ in zero
external magnetic field.\cite{Movshovich} Balatsky\cite{Balatsky}
pointed out that, in the presence of magnetic impurities, the
$d^{}_{x^2-y^2}$ superconductor can exhibit a transition exactly
into the $d+id$ state due to a coupling between impurity
magnetization and the $d^{}_{xy}$ component of the order
parameter.

In this Communication, we report a possibility of a rise of $d+id$
chiral state in high-temperature SC cuprates.

Recently introduced concept of SC pairing with large pair momentum
under screened Coulomb repulsion\cite{BK_06} offers an explanation
of principal features of cuprate superconductors: 1) a
checkerboard real-space ordering observable in the SC
state\cite{Wise} can be directly related to the pair momentum,
comparable with reciprocal lattice spacing;\cite{BK_09} 2)
pseudogap (PG) state\cite{NPK} with a broad region of SC
fluctuations above the transition temperature $T^{}_c$ can be
explained by a rise of quasi-stationary states of pairs with large
momentum due to real-space oscillations of the screened Coulomb
potential;\cite{BK_07} 3) high-energy effects observable in
optical experiments\cite{Basov_Timusk} can be related to the
electron-hole asymmetry that becomes apparent in the SC state of
the cuprates.

The order parameter originating from SC pairing with large
momentum is nonvanishing in the interior of a part of the
Brillouin zone (domain of kinematic constraint) due to the fact
that, at $T=0$, the momenta of both particles composing a pair
should be either inside or outside the Fermi contour (FC). This
order parameter turns out to be appreciably nonzero inside
vicinities of nested segments of the FC.\cite{BK_07} In the case
of SC cuprates, such segments correspond to antinodal region of
the Brillouin zone.\cite{Wise}

High values of $T^{}_c$ and specific isotope effect manifested in
the cuprates\cite{Gweon} show that, together with the repulsive
Coulomb pairing interaction, one should take into account the
attractive contribution owing to electron-phonon interaction
(EPI), including the forward scattering effect.\cite{Dolgov}

In the antinodal region, phonon assisted Coulomb pairing with
large momentum can predominate over conventional phonon-induced
pairing with zero momentum that prevails only in the nodal region.
Then, superconductivity at low temperatures should exist as a
biordered state formed by the condensates of pairs with large and
zero momenta in the antinodal and nodal regions of the momentum
space, respectively.\cite{BK_07} On the contrary, the SC order
just below $T^{}_c$ should be determined by the pairing with large
momentum. Therefore, such an order should arise only in the
antinodal region.

Considerable enhancement of $T^{}_c$, observable in the cuprates,
can be qualitatively related to specific phonon induced
``symmetrization'' of real-space oscillations of the screened
Coulomb potential.\cite{BKKM} The SC order parameter
${\Delta}({\bm{k}};{\bm{K}})$, where ${\bm{k}}$ and ${\bm{K}}$ are
relative motion and center-of-mass momenta of SC pair,
respectively, should be obtained as a self-consistent solution to
the mean-field gap equation with the momentum representation of
such a symmetrized potential
$U({\bm{k}},{\bm{k}}_{}^{\prime};{\bm{K}})$. The gap equation at
$T=0$ can be written as
\begin{equation}\label{OP}
{\Delta}({\bm{k}};{\bm{K}})=-{\frac{1}{2}}\sum\limits
_{{\bm{k}}_{}^{\prime}}
{\frac{U({\bm{k}},{\bm{k}}_{}^{\prime};{\bm{K}})\,
{\Delta}({\bm{k}}_{}^{\prime};{\bm{K}})}{{\sqrt{{\epsilon}^2_+
({\bm{k}}_{}^{\prime};{\bm{K}})+
{\Delta}_{}^2({\bm{k}}_{}^{\prime};{\bm{K}})}}}}.
\end{equation}
Here, $2{\epsilon}^{}_+
({\bm{k}};{\bm{K}})={\varepsilon}({\bm{k}}^{}_+)
+{\varepsilon}({\bm{k}}^{}_-)$ is kinetic energy of a singlet pair
composed of the particles with momenta
${\bm{k}}^{}_{\pm}={\bm{K}}/2\pm{\bm{k}}$, the summation is taken
over momenta ${\bm{k}}^{\prime}_{}$ belonging to the domain of
kinematic constraint relevant to given pair momentum ${\bm{K}}$.

In the case of pairing with ${\bm{K}}\neq 0$, in common with the
well-known Fulde-Ferrel-Larkin-Ovchinnikov (FFLO)
problem,\cite{FF,LO} the order parameter can be represented in the
form of either a running wave,\cite{FF} ${\Delta}\sim
{\exp{i{\bm{KR}}}}$, or a standing wave,\cite{LO} ${\Delta}\sim
{\cos{({\bm{KR}})}}$, that is as a symmetric superposition of
running waves with pair momenta $\pm{\bm{K}}$. Here, ${\bm{R}}$ is
center-of-mass radius-vector of the pair. It should be noted that,
unlike the FFLO state, the SC state with nonzero center-of-mass
momentum considered here arises without external magnetic field
and, generally speaking, preserves TR symmetry.

One can expect that, along with the symmetric superposition,
antisymmetric superposition of the same waves, ${\Delta}\sim
{\sin{({\bm{KR}})}}$, could be a solution to the gap equation as
well. Both symmetric and antisymmetric solutions are defined in
common domain of kinematic constraint which has to be constructed
as the union of the domains for running waves with $\pm{\bm{K}}$.

Because of the crystal symmetry of the system, FFLO order
parameter can be defined as a more complicated linear combination
of running waves with equivalent momenta.\cite{Matsuda} In a
similar way, one can define zeroth-order approximation of the
order parameter arising as a result of SC pairing with large pair
momentum.

In the case of the cuprates, tetragonal symmetry of
${\text{CuO}}_2$ plane results in four crystal equivalent pair
momenta: $\pm{\bm{K}}$ and $\pm{\bm{K}}_{}^{\prime}$ where
${\bm{K}}_{}^{\prime}$ is perpendicular to $\pm{\bm{K}}$. It is
convenient to form standing waves as symmetric and antisymmetric
(with respect to in-plane reflection from a line perpendicular to
pair momentum) superpositions for each of two running waves with
momenta, $\pm{\bm{K}}$ and $\pm{\bm{K}}_{}^{\prime}$,
respectively. Then, the order parameter in the whole of the
Brillouin zone can be written as a linear combination of these
standing waves. Coefficients in such linear combinations specifies
the orbital symmetry of the order parameter.\cite{BKNT}

Coefficients of like signs correspond to extended $s$-wave
symmetry (the order parameter is invariant with respect to
rotation by ${\pi}/2$ about $C^{}_4$ axis). In the case of
coefficients of unlike signs, the order parameter reverses sign
under rotation by ${\pi}/2$ and therefore can be referred to
$d$-wave orbital symmetry.

One can represent the order parameter by any of four linear
combinations (two $s$-wave and two $d$-wave) directly following
from the gap equation. The SC ground state of the system should be
expressed by the linear combination which has a lower free energy.

For solving the gap equation, we present interaction energy
$U({\bm{k}},{\bm{k}}_{}^{\prime};{\bm{K}})$ as a sum of screened
Coulomb repulsion, $U^{}_s({\bm{k}},{\bm{k}}_{}^{\prime})$,
defined in the whole of the domain of kinematic constraint and EPI
induced effective attraction which is assumed nonzero inside a
narrow region enveloping the FC.\cite{BK_07} Width of this region
in the momentum space is of the order of $2{\omega}^{}_D/v^{}_F$
where ${\omega}^{}_D$ and $v^{}_F$ are characteristic Debye
frequency and Fermi velocity normal to the FC, respectively. We
assume that attractive contribution into interaction energy is
nonzero if and only if momenta of particles before and after
scattering (${\bm{k}}$ and ${\bm{k}}_{}^{\prime}$, respectively)
both belong to this region. Also, we assume that this contribution
is independent of momenta inside the region.

Thus, $U({\bm{k}},{\bm{k}}_{}^{\prime};{\bm{K}})=
U^{}_s({\bm{k}},{\bm{k}}_{}^{\prime})-V$, if both momenta belong
to the region, $U({\bm{k}},{\bm{k}}_{}^{\prime};{\bm{K}})=
U^{}_s({\bm{k}},{\bm{k}}_{}^{\prime})$ when even if one of the
momenta ${\bm{k}}$ and ${\bm{k}}_{}^{\prime}$ belonging to the
domain of kinematic constraint does not belong to the region. One
can assume that $U({\bm{k}},{\bm{k}}_{}^{\prime};{\bm{K}})=0$ when
even if one of the momenta does not belong to the domain of
kinematic constraint.

To solve the gap equation numerically, we approximate the
interaction energy by a simple step function of
${\bm{\kappa}}={\bm{k}}-{\bm{k}}_{}^{\prime}$.\cite{BK_09} Step
length in the direction of ${\bm{K}}$ is limited by the length of
the nested segment of the FC. In the direction perpendicular to
${\bm{K}}$, left step at ${\kappa}^{}_l$ corresponds to
phonon-mediated decrease in energy taking into account the forward
scattering effect\cite{Dolgov} whereas right step at
${\kappa}^{}_r$ reflects the fact that values of the order
parameter turn out to be very small at all points distant from the
FC. Therefore, the order parameter turns out to be weakly
sensitive to ${\kappa}^{}_r$.\cite{BK_09}

We use the electron dispersion that conforms to the FC observable
in hole doped cuprates,
\begin{eqnarray}\label{DL}
{\varepsilon}(k^{}_x,k^{}_y)&=&
2t^{}_0-2t({\cos}k^{}_x+{\cos}k^{}_y)
-4t_{}^{\prime}{\cos}k^{}_x{\cos}k^{}_y-\nonumber \\ &-&
2t_{}^{\prime \prime}({\cos}2k^{}_x+{\cos}2k^{}_y),
\end{eqnarray}
where $t^{}_0=1\, {\text{eV}}$, $t=0.5\, {\text{eV}}$,
$t_{}^{\prime}/t=-0.3$, $t_{}^{\prime \prime}/t=0.14$, and
dimensionless components of momentum $k^{}_i$ ($i=x,y$) vary
within $-{\pi}<k^{}_i \leq {\pi}$.

\begin{figure}
\includegraphics[scale=.54]{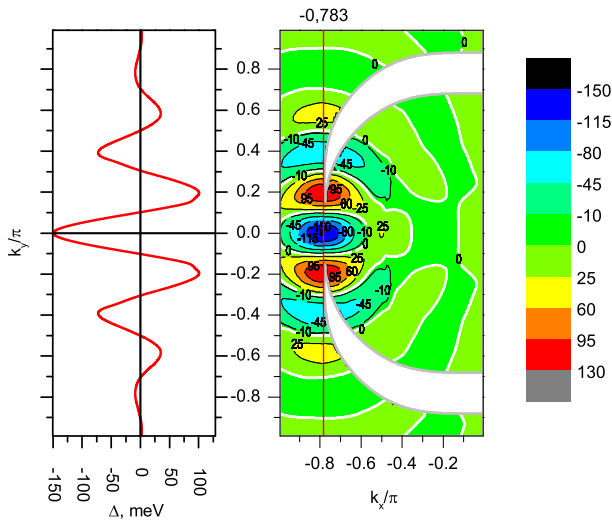}
\caption[*]{Symmetric superposition
${\Delta}^{}_s({\bm{k}};{\bm{K}})$ of running waves with wave
vectors $\pm{\bm{K}}$ directed along $k^{}_y$ axis. Left panel:
dependence of ${\Delta}^{}_s({\bm{k}};{\bm{K}})$ on $k^{}_y$ at
$k^{}_x=-0.783{\pi}$ that corresponds to the position of the left
vertical nested segment of the FC. Right panel: topology of
${\Delta}^{}_s({\bm{k}};{\bm{K}})$; only left half of the
Brillouin zone is shown. White lines present the intrinsic zero
lines on which ${\Delta}^{}_s({\bm{k}};{\bm{K}})=0$. White
background between the Fermi contours shifted by $\pm{\bm{K}}/2$
with respect to initial position of the FC reflects the kinematic
constraint.} \label{sym_c.eps}
\end{figure}

Recently obtained numerical solution to the gap
equation\cite{BK_09} corresponding to a symmetric superposition of
two running waves is presented in Fig.~1 (only left half of the
Brillouin zone is shown) in which the momenta of the running waves
$\pm{\bm{K}}$ are chosen as directed along $k^{}_y$ axis. The
order parameter ${\Delta}^{}_s({\bm{k}};{\bm{K}})$ is
characterized by intrinsic system of zero lines intersecting the
FC. One can see that ${\Delta}^{}_s({\bm{k}};{\bm{K}})$ possesses
distinct values only in a vicinity of nested segments of the FC.
Similarly, one can obtain the symmetric superposition
${\Delta}^{}_s({\bm{k}};{\bm{K}}_{}^{\prime})$ corresponding to
the running waves with $\pm{\bm{K}}_{}^{\prime}$.

The symmetric superposition ${\Delta}^{}_s({\bm{k}};{\bm{K}})$ can
describe a stripe structure of the SC state,\cite{BK_09} in
particular, an emergence of the SC order parameter appearing in
the PG state of the cuprates.\cite{Wise} Recently, Berg et
al.\cite{Berg} discussed similar striped SC state as a
unidirectional pair-density wave (PDW) phase with periodic
real-space dependence of the order parameter on the center-of-mass
position. The coupling between such a PDW and other ordered states
was also considered in the framework of Ginzburg-Landau
theory.\cite{Berg}

To obtain the order parameter in the whole of the Brillouin zone,
one should compose either $s$-wave or $d$-wave linear combination
of obtained symmetric superpositions,\cite{BK_06}
\begin{equation}\label{sym}
{\Delta}_s^{(\pm)}({\bm{k}})\sim
{\Delta}^{}_s({\bm{k}};{\bm{K}})\pm
{\Delta}^{}_s({\bm{k}};{\bm{K}}_{}^{\prime}).
\end{equation}
The first of them (${\Delta}_s^{(+)}$), corresponding to extended
$s$-wave orbital symmetry, displays the intrinsic zero lines only,
whereas the second one (${\Delta}_s^{(-)}$), corresponding to
$d^{}_{x^2-y^2}$ orbital symmetry, besides the intrinsic zero
lines, displays four straight zero lines (nodal lines) along the
diagonals of the Brillouin zone.

In the cuprates, it seems that such orbital $d$-wave nodal lines
are consistent with available experimental facts including
angle-resolved photoemission spectroscopy (ARPES)
data.\cite{Damascelli}. In particular, it is very likely that SC
gap near the diagonals takes small values and can even vanish. On
the contrary, intrinsic zero lines, situated close to the FC,
hardly ever can be detected directly from ARPES measurements but
they undoubtedly should become apparent in thermodynamical
properties.

In Fig.~2, we present new numerical solution to the gap equation,
namely, the order parameter in the form of antisymmetric
superposition of two running waves with opposite momenta,
${\Delta}^{}_a({\bm{k}};{\bm{K}})$.

\begin{figure}
\includegraphics[scale=.54]{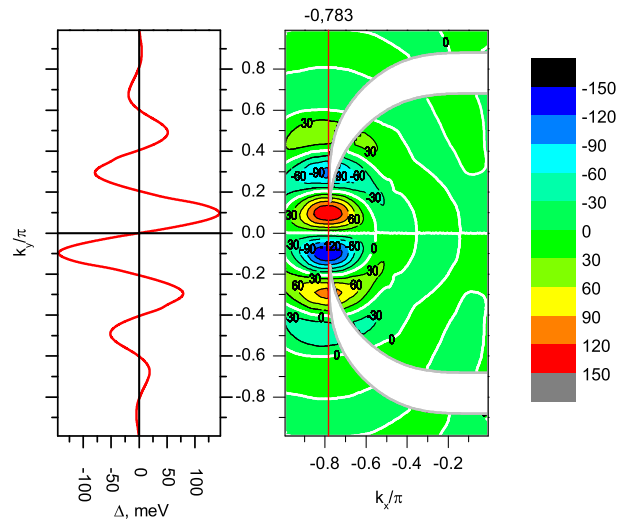}
\caption[*]{Antisymmetric superposition
${\Delta}^{}_a({\bm{k}};{\bm{K}})$ of running waves with wave
vectors $\pm{\bm{K}}$ directed along $k^{}_y$ axis. Left panel:
dependence of ${\Delta}^{}_a({\bm{k}};{\bm{K}})$ on $k^{}_y$ at
$k^{}_x=-0.783{\pi}$ that corresponds to the position of the left
vertical nested segment of the FC. Right panel: topology of
${\Delta}^{}_a({\bm{k}};{\bm{K}})$; only left half of the
Brillouin zone is shown. White lines present the intrinsic zero
lines on which ${\Delta}^{}_a({\bm{k}};{\bm{K}})=0$. White
background between the Fermi contours shifted by $\pm{\bm{K}}/2$
with respect to initial position of the FC reflects the kinematic
constraint.} \label{asym_c1.eps}
\end{figure}

This solution, just as symmetric superposition
${\Delta}^{}_s({\bm{k}};{\bm{K}})$, has its own system of
intrinsic zero lines. As one can see from Fig.~2, one of such
lines turns out to be parallel to one of the coordinate axes.

Extreme values of both ${\Delta}^{}_s({\bm{k}};{\bm{K}})$ and
${\Delta}^{}_a({\bm{k}};{\bm{K}})$ are concentrated in the
antinodal regions near nested segments of the FC. It should be
noted especially that, as results from direct numerical solution
to the gap equation, both symmetric and antisymmetric
superpositions exhibit comparable extreme values.

The order parameter in the whole of the Brillouin zone should be
presented as either $s$-wave or $d$-wave linear combination of
antisymmetric superpositions,
\begin{equation}\label{anti}
{\Delta}_a^{(\pm)}({\bm{k}})\sim
{\Delta}^{}_a({\bm{k}};{\bm{K}})\pm
{\Delta}^{}_a({\bm{k}};{\bm{K}}_{}^{\prime}).
\end{equation}
The $s$-wave combination corresponding to plus sign in
Eq.~(\ref{anti}), besides some closed intrinsic zero lines,
displays eight straight zero lines parallel both sides and
diagonals of the Brillouin zone. Formally, such an order parameter
can be referred to the so-called $g$-wave orbital symmetry
discussed by Zhao.\cite{Zhao}

The $d$-wave combination corresponding to minus sign in
Eq.~(\ref{anti}), besides some closed intrinsic zero lines,
displays four intrinsic straight zero lines directed along the
sides of the Brillouin zone. These lines can be formally
considered as the nodes of the order parameter with $d^{}_{xy}$
orbital symmetry.

It is the pairing interaction that makes the choice in favor of a
certain set of coefficients in Exs.~(\ref{sym}) and (\ref{anti}).
For example, near a spin-density-wave instability typical of the
cuprate superconductors, paramagnon exchange induced component of
the pairing interaction, sensitive to band structure and band
filling, gives rise to singlet $d$-wave pairing with zero
center-of-mass momentum. Such an interaction favors (suppresses)
$d^{}_{x^2-y^2}$ ($d^{}_{xy}$) channel.\cite{Scalapino}

As follows from Figures~1 and 2, the amplitudes of two
superpositions, ${\Delta}^{}_s({\bm{k}};{\bm{K}})$ and
${\Delta}^{}_a({\bm{k}};{\bm{K}})$, are comparable. Therefore,
under certain conditions, the order parameter with spontaneously
broken TR symmetry,
\begin{equation}\label{d+id}
{\Delta}_{}^{(c)}({\bm{k}})\sim {\Delta}^{(-)}_s({\bm{k}})+i
{\Delta}^{(-)}_a({\bm{k}}),
\end{equation}
can be expected as a chiral fully gapped ground state of the
system. The components of ${\Delta}_{}^{(c)}({\bm{k}})$, phased by
${\pi}/2$ with respect to each other, should be formally referred
to $d$-wave orbital symmetry, $d^{}_{x^2-y^2}$ and $d^{}_{xy}$,
respectively.

It should be noted that, in the problem of triplet pairing with
large center-of-mass momentum, superpositions
${\Delta}^{}_a({\bm{k}};{\bm{K}}_{}^{\prime})$ and
${\Delta}^{}_a({\bm{k}};{\bm{K}})$ themselves, can be considered
as the $p$-wave order parameters corresponding to $p^{}_x$ and
$p^{}_y$ symmetry, respectively. These superpositions can be used
to form chiral triplet SC state with $p^{}_x+ip^{}_y$ orbital
symmetry.

Values of the $d^{}_{xy}$ component, ${\Delta}^{(-)}_a({\bm{k}})$,
should be nonzero for momenta belonging to the nodal region
because of the proximity effect in the momentum space and a
contribution of SC pairing with zero momentum into biordered SC
state\cite{BK_07} (both ignored in Figures~1 and 2). As a result,
a finite SC gap should appear at the points that correspond to the
nodes of the pure $d^{}_{x^2-y^2}$ state.

The $d^{}_{xy}$ order parameter is often accepted in a simple
form, ${\sin}k^{}_x {\sin}k^{}_y$, that reveals maxima just on the
diagonals. On the contrary, $d+id$ order (\ref{d+id}) turns out be
concentrated in antinodal vicinities of the nested segments of the
FC (see Fig.~2). Therefore, the smallness of the $d^{}_{xy}$
component of Ex.(\ref{d+id}) in the nodal region can make it
difficult to detect such a gap (for example, using ARPES
technique). It should be noted, however, that recent ARPES
data\cite{Okawa} can be considered as an unambiguous evidence in
favor to $d^{}_{x^2-y^2}$-wave-like SC gap in optimally doped
${\text{YBa}}_2{\text{Cu}}_3{\text{O}}_{7-\delta}$ that exhibits a
nonzero minimum of about $12\, meV$ along the nodal direction.

Chiral SC ground state with the order parameter (\ref{d+id}) turns
out to be topologically nontrivial. Indeed, by continuous
deformation of the parameters of the mean-field Hamiltonian
without opening a gap, this state can be transformed into that
considered by Volovik:\cite{Volovik_1} both states are
topologically equivalent and should be characterized by the
topological invariant ${\cal{N}}=\pm 2$.


\begin{acknowledgements}

This work was supported in part by the Russian Foundation for
Basic Research (Project 11-02-01149-a), The Ministry of Education
and Science of the Russian Federation (contract 16.513.11.3149),
and Presidium of Russian Academy of Sciences (Program of Basic
Research No.~24, Project 1.1.6.3).

\end{acknowledgements}

\end{document}